\documentclass[11pt]{article}

\usepackage[T1]{fontenc}
\usepackage[utf8]{inputenc}
\usepackage{lmodern}
\usepackage{microtype}
\usepackage[margin=1.08in]{geometry}
\usepackage{amsmath,amssymb,amsthm,mathtools}
\allowdisplaybreaks
\usepackage{enumitem}
\usepackage[colorlinks=true,linkcolor=blue,citecolor=blue,urlcolor=blue]{hyperref}
\usepackage{newpxtext}
\usepackage{newpxmath}
\newtheorem{theorem}{Theorem}[section]
\newtheorem{proposition}[theorem]{Proposition}

\newtheorem{corollary}[theorem]{Corollary}

\theoremstyle{definition}
\newtheorem{definition}[theorem]{Definition}
\theoremstyle{remark}

\let\emptyset\varnothing

\newcommand{\cp}[0]{\{0, 1\}^\infty}
\newcommand{\fs}[0]{\{0, 1\}^{*}}

\title{Solomonoff Induction with Reflective Oracles\\
{\small PIBBSS Fellowship Project}}
\author{
    Damini Kusum\\[0.5em]
    {Mentors: Cole Wyeth and Aram Ebtekar (AIXI Labs)}
}
\date{August 2026}

\begin{document}
\maketitle

\section{Introduction}
\label{sec:introduction}

Universal induction can be viewed as competitive prediction against effective predictive strategies. In the tradition of Solomonoff induction \cite{Solomonoff1964a,Solomonoff1964b,Hutter2005,LiVitanyi2008}, this can be formulated as a recommendation regarding the priors of an ideal Bayesian predictor. A hypothesis space \(\mathcal C\) consisting of probabilistic data-generating sources\footnote{We will restrict attention to infinite binary sequences, but the results in this document can be extended to data generated with finite alphabet.} is fixed, where each hypothesis receives prior weight decreasing with the length of its shortest description. The predictor \(\xi\), obtained by mixing the class according to these weights, makes next-bit predictions by conditionalizing on the observed history. Importantly, \(\xi\) enjoys the performance property of optimality, that is, \(\xi\) multiplicatively dominates every \(\mu\in\mathcal C\). Dominance guarantees that no hypothesis in \(\mathcal C\) is more than boundedly better supported by the accumulating evidence than the predictor itself. \\

Taking \(\mathcal C\) to consist of all computable probability measures would seem a reasonable choice. The difficulty, however, is that the class of computable measures does not admit an effective enumeration. Solomonoff induction remedies this by expanding the hypothesis space to include lower semicomputable semimeasures. Semimeasures can be thought of as defective measures, which may lose probability mass as a string is extended. While the effective enumerability of the class of lower semicomputable semimeasures makes it possible for the class to contain an optimal member, this optimal member is inevitably only lower semicomputable. Its conditional probabilities are, in general, only limit computable. This obstruction persists relative to any deterministic oracle $A$. An optimal lower $A$-semicomputable semimeasure has conditional probabilities that are limit computable relative to $A$, but not $A$-computable. Passing to the Turing jump $A'$ makes these conditional probabilities computable, but those of an optimal predictor over the enlarged class of lower $A'$-semicomputable semimeasures are again not $A'$-computable. \\

Reflective oracles address this mismatch within a probabilistic model of computation. A Turing machine may ask probabilistic questions about the output of machines that use the same oracle, and the oracle may randomize at boundary cases where deterministic answers would generate a diagonal contradiction \cite{FallensteinTaylorChristiano2015}. For a fixed reflective oracle \(O\), completion turns probabilistic Turing machines with oracle access into proper probability measures. The resulting class \(\mathcal M^O_{\mathrm{refl}}\) is effectively indexed by oracle machines, its completed conditionals are uniformly computable in \(O\), and it contains a member that multiplicatively dominates the entire class \cite{FallensteinSoaresTaylor2015,LeikeTaylorFallenstein2016,WyethEtAl2025}. Thus reflective computation provides an optimal predictor \(\xi^O\)  whose forecasts can themselves be used to construct further hypotheses within the class.\\

In this document, we develop a purely inductive consequence of this closure by addressing the following question: can the optimal predictor for the reflective variant of Solomonoff induction compete with an effective predictive strategy that specifies only part of a data-generating process? A strategy may recognize, for example, that every even bit repeats the preceding bit, while having nothing useful to say about the odd bits. The unrestricted bits need not be generated computably, so the usual convergence guarantees under a computable source do not directly address this problem. Lattimore, Hutter, and Gavane \cite{LattimoreHutterGavane2011} show that the ordinary Solomonoff semimeasure can fail to learn even this simple regularity, although its normalized version is shown to learn all such computable sub-patterns.\\

This problem is naturally expressed in terms of incomplete models. An incomplete model specifies some features of a data-generating process while leaving others unresolved, and is, therefore, compatible with a family of probability laws. A model may, for example, computably specify the conditional distribution of a label given its input while remaining silent about the process generating the inputs. To compare $\xi^O$ with such a partial model, we construct a comparison hypothesis that follows the model at histories where it makes a prediction and uses $\xi^O$’s conditional probabilities at all other histories. Reflective closure places these hypotheses in the class dominated by $\xi^O$, and dominance bounds their cumulative log-loss advantage in terms of description length. More generally, when a model specifies computable lower and upper bounds on next-bit probabilities, $\xi^O$'s predictions approach the permitted intervals almost surely under every compatible probability law satisfying those bounds. Deterministic and stochastic selected-bits guarantees follow as special cases in the reflective setting. The same comparison argument also yields weighted unbiasedness and a corresponding calibration property on every sequence.\\

Finally, we ask whether the probabilities assigned by an optimal predictor suffice to compute every hypothesis in its class. Our comparison arguments use the fact that hypotheses computable from the predictions of $\xi^O$ remain within $\mathcal M^O_{\mathrm{refl}}$. The converse is a separate question: in general, multiplicative dominance does not imply that the dominated hypothesis is computable from the predictor. We construct an optimal predictor $\widehat{\xi}^O$ for which both directions hold: the reflective class consists exactly of the hypotheses computable from its predictions.\\

This document is organized as follows. The remainder of this section introduces our notation and reviews the relevant background on reflective oracles. Section~2 establishes closure under forecast-dependent constructions, derives a predictor-relative bound based on the description length, and proves results on incomplete models, prediction of selected bits, and calibration. Section~3 establishes the representation result for an optimal predictor presenting the reflective hypothesis class. The appendix extends the incomplete-model result to finite-horizon prediction.

\subsection{Preliminaries and Notation}

\textit{Strings and Sequences.} Let \(\cp\) and \(\fs\) denote the sets of infinite binary sequences and finite binary strings, respectively. Given a sequence \(\alpha\in\cp\), we use \(\alpha_n\) to denote its \(n\)-th bit, \(\alpha_{1:n}\) to denote its first \(n\) bits, and \(\alpha_{<n}\) to denote \(\alpha_{1:n-1}\). The empty string is denoted by \(\epsilon\), and the length of a finite string \(\sigma\) by \(|\sigma|\). We restrict our attention to Cantor space, which is topologized by the collection of cylinder sets of the form \([\sigma]\), where \(\sigma\in\fs\) and \([\sigma]=\{\alpha\in\cp:\sigma\prec\alpha\}\). Here, \(\sigma\prec\alpha\) means that \(\sigma\) is a prefix of \(\alpha\), or equivalently, that \(\alpha\) is an extension of \(\sigma\). For \(A\subseteq\fs\), let \([A]=\bigcup_{\sigma\in A}[\sigma]\). \\

\textit{Measures and Semimeasures.} For some data-generating process \(\mu\), the probability space in which we work is given by the triple \((\cp,\mathcal{B}(\cp),\mu)\), where \(\mathcal{B}(\cp)\) is the Borel \(\sigma\)-algebra on Cantor space, that is, the smallest \(\sigma\)-algebra containing all basic open sets. Any function \(f:\fs\rightarrow[0,1]\) satisfying \(f(\epsilon)=1\) and \(f(\sigma)=f(\sigma0)+f(\sigma1)\) for every \(\sigma\in\fs\) determines a unique Borel probability measure on \(\cp\). Conversely, every Borel probability measure \(\mu\) on \(\cp\) is uniquely determined by its values on the basic open sets. For convenience, we write \(\mu(\sigma)\) for \(\mu([\sigma])\). \(\mu\) is said to have full support when \(\mu(\sigma)>0\) for every \(\sigma\in\fs\). Whenever \(\mu(\sigma)>0\), we write \(\mu(a|\sigma)=\mu(\sigma a)/\mu(\sigma)\) for \(a\in\{0,1\}\). A semimeasure is a function \(\nu:\fs\rightarrow[0,1]\) satisfying \(\nu(\epsilon)\leq 1\) and \(\nu(\sigma)\geq\nu(\sigma0)+\nu(\sigma1)\) for every \(\sigma\in\fs\). Thus, unlike a probability measure, a semimeasure may lose probability mass as a string is extended. When \(\nu(\sigma)>0\), we similarly define \(\nu(a|\sigma)=\nu(\sigma a)/\nu(\sigma)\).\\

\textit{Computability.} A real $r\in\mathbb{R}$ is computable if it has a computable Cauchy name, where a Cauchy name for $r$ is a function $\varphi:\mathbb{N}\rightarrow\mathbb{Q}$ satisfying $|\varphi(n)-r|\leq 2^{-n}$ for every $n$. A real is left c.e.\ if it is the limit of a computable nondecreasing sequence of rationals, and limit computable if it is the limit of a computable sequence of rationals, with no computable rate of convergence required. These notions extend to real-valued functions on discrete domains by requiring the approximations to be uniform in the input. For functions, we use the term lower semicomputable for the analogue of left c.e.\\

\textit{Kolmogorov Complexity.} A set $P\subseteq\fs$ is prefix-free if no element of $P$ is a proper prefix of another. Fix a universal prefix-free Turing machine $U$, whose halting domain is prefix-free and which simulates every other prefix-free Turing machine with an additive overhead in program length depending only on the simulated machine. The prefix-free Kolmogorov complexity of a finite string \(x\) is \(K(x)=\min\{\,|p|:U(p)=x\,\}\), where \(p\) ranges over finite binary programs. More generally, if a mathematical object \(o\) is represented within a fixed effective coding scheme, \(K(o)\) denotes the length of its shortest description under that scheme. In particular, given an effective enumeration \((d_i)_{i\in\mathbb{N}}\), we may define \(K(d)=\min\{K(i):d_i=d\}\). For a fixed representation of the objects, changing the universal prefix machine changes \(K\) by at most an additive constant.\\

\textit{Probabilistic Turing Machines.} A probabilistic Turing machine (pTM) is a Turing machine equipped with an additional tape of independent fair random bits. We regard a pTM $T$ as a sampling procedure for the next bit given a finite history $\sigma$. Let $\lambda_T(a|\sigma)$ denote the probability that $T$, on input $\sigma$, halts and outputs $a\in\{0,1\}$. Since $T$ need not halt, we may have $\lambda_T(0|\sigma)+\lambda_T(1|\sigma)<1$. Repeatedly applying this procedure with fresh randomness assigns probabilities to finite prefixes, given by $\lambda_T(\epsilon)=1$ and $\lambda_T(\sigma a)=\lambda_T(\sigma)\lambda_T(a|\sigma)$. These assignments define a semimeasure on $\cp$. If $T$ halts almost surely at every history, they determine a probability measure on $\cp$. The output probabilities $\lambda_T(a|\sigma)$, and hence the induced semimeasure $\lambda_T$, are uniformly lower semicomputable.\\

\textit{Reflective Oracles.} A reflective oracle allows probabilistic Turing machines to query the behavior of other probabilistic Turing machines that have access to the same oracle. We write \(T^O\) for a pTM \(T\) with access to \(O\), and let \(\lambda_T^O(a|\sigma)\) denote the probability that \(T^O\), on input \(\sigma\), halts and outputs \(a\). As above, these conditional probabilities induce a semimeasure \(\lambda_T^O\).\\

A query to \(O\) consists of a pTM \(T\), a history \(\sigma\in\fs\), and a rational threshold \(p\in[0,1]\). We identify \(O(T,\sigma,p)\in[0,1]\) with the probability that the oracle answers \(1\) to this query. Each oracle call uses fresh randomness. The oracle \(O\) is \emph{reflective}\footnote{We use the step reflective oracle convention of \cite{WyethEtAl2025}.} if, for every \(T\) and \(\sigma\), there is a value \(q_{T,\sigma}^O\in[0,1]\) such that \(\lambda_T^O(1|\sigma)\leq q_{T,\sigma}^O\leq 1-\lambda_T^O(0|\sigma)\), and for every rational \(p\), \(O(T,\sigma,p)=1\) whenever \(p<q_{T,\sigma}^O\), while \(O(T,\sigma,p)=0\) whenever \(p>q_{T,\sigma}^O\). At the boundary \(p=q_{T,\sigma}^O\), the oracle may randomize. If \(T^O\) halts almost surely with binary output, then \(\lambda_T^O(0|\sigma)+\lambda_T^O(1|\sigma)=1\), so \(q_{T,\sigma}^O=\lambda_T^O(1|\sigma)\). In this case the oracle simply answers whether the probability that \(T^O\) outputs \(1\) is greater or smaller than \(p\), with randomization permitted at equality. Reflective oracles exist \cite{FallensteinTaylorChristiano2015}; moreover, a reflective oracle can be chosen to be limit computable \cite{LeikeTaylorFallenstein2016,WyethEtAl2025}. The value \(q_{T,\sigma}^O\) given by reflectivity also specifies how any probability mass lost through nonhalting is to be distributed. Define the completed conditional probabilities by \(\bar{\lambda}_T^O(1|\sigma)=q_{T,\sigma}^O\) and \(\bar{\lambda}_T^O(0|\sigma)=1-q_{T,\sigma}^O\). Since these probabilities sum to one at every history, setting \(\bar{\lambda}_T^O(\epsilon)=1\) and extending by the chain rule defines a proper probability measure on \(\cp\). We refer to \(\bar{\lambda}_T^O\) as the completion of \(T\) with respect to \(O\). The completed conditional probabilities are uniformly computable with access to \(O\). Moreover, any family of binary conditional probabilities that sums to one at every history and is uniformly computable with access to \(O\) can be sampled by a probabilistic Turing machine with access to \(O\) that halts almost surely at every history \cite{WyethEtAl2025}.

\section{Reflective Prediction and Comparison Hypotheses}
\label{sec:reflective-comparison}

The predictive results in this section use a common comparison argument. We construct a hypothesis by a computable modification of the predictor's conditional probabilities, show that it belongs to the reflective class, and apply dominance to bound its cumulative log-loss advantage. We first establish the closure and comparison results, then apply them to incomplete models, selected bits, and calibration. In Section~3, we establish an existential converse to closure: an optimal predictor can be chosen so that every member of the reflective class is computable from its predictions.

\subsection{The reflective hypothesis class}
\label{subsec:reflective-class}

Fix a reflective oracle $O$. Let $\mathcal M_{\mathrm{refl}}^O=\{\bar{\lambda}_T^O:T\text{ is a probabilistic Turing machine with access to }O\}$ be the class of measures induced by completing the output probabilities of these machines with respect to $O$. An effective enumeration $\{T_e\}_{e\in\mathbb N}$ of probabilistic Turing machines with oracle access gives an indexing $\mu_e=\bar{\lambda}_{T_e}^O$ of this class. We use the construction of Wyeth et al.\ \cite[Theorem~20]{WyethEtAl2025} to obtain an optimal predictor.\\

In the spirit of Solomonoff induction, assign each completed measure $\nu\in\mathcal M_{\mathrm{refl}}^O$ a positive weight $w_\nu$ satisfying $\sum_\nu w_\nu\leq1$, and consider the weighted sum $\sum_\nu w_\nu\nu$. For our purposes, we take $w_\nu=2^{-K(\nu)}$. Kraft's inequality ensures that $\sum_\nu2^{-K(\nu)}\leq1$, and the resulting weighted sum is uniformly lower semicomputable with access to $O$.\footnote{This sum is the supremum of $\sum_{e\in F}2^{-K(e)}\mu_e(\sigma)$ over finite sets $F$ whose completed measures are pairwise distinct. Such sets can be enumerated with access to $O$ by finding cylinders on which each pair of measures differs. Since $K(e)$ is uniformly approximable from above and $\mu_e(\sigma)$ is uniformly computable with access to $O$, these finite sums are uniformly lower semicomputable with access to $O$.} Their algorithm defines a probabilistic Turing machine with access to $O$ whose completed measure $\xi^O\in\mathcal M_{\mathrm{refl}}^O$ satisfies $\xi^O(\sigma)\geq\sum_\nu w_\nu\nu(\sigma)$ for every $\sigma\in\fs$. Thus $\xi^O$ belongs to the reflective class and multiplicatively dominates every member of it. We fix this predictor throughout.\\

The class contains the fair-coin measure\footnote{The fair-coin measure $\lambda$, also known as the uniform measure, is determined by $\lambda(\sigma)=2^{-|\sigma|}$ for every $\sigma\in\fs$.}, so dominance ensures that $\xi^O$ has full support. Its completed conditionals are uniformly computable with access to $O$ and, by full support, coincide with its conditional probabilities at every history. A probabilistic Turing machine with access to $O$ can therefore approximate the forecasts of $\xi^O$ to any prescribed accuracy and use these approximations to determine its own output probabilities. In the next subsection, we show that every hypothesis computable from the predictor's forecasts belongs to $\mathcal M_{\mathrm{refl}}^O$.
\subsection{Constructing hypotheses from forecasts}
\label{subsec:forecast-constructions}

We first specify what it means for a hypothesis to be computable from a predictor's forecasts.

\begin{definition} A conditional presentation of a probability measure $\mu$ is a total function
$p_\mu:\{0,1\}^*\to[0,1]$
such that $p_\mu(\sigma)=\mu(1|\sigma)$ whenever $\mu(\sigma)>0.$ At histories with $\mu$-probability zero, the values of $p_\mu$ may be chosen arbitrarily. Let $\operatorname{Name}(\mu)$ be the set of names for all conditional presentations of $\mu$.
\end{definition}

A full-support measure has a unique conditional presentation, although that presentation has many names.

\begin{definition}
Let $\rho$ be a probability measure. We write
$\mu\preceq_{\mathrm{pred}}\rho$ if there exists a Turing machine $T$ with oracle access such that, for every
$\varphi\in\operatorname{Name}(\rho)$, the function
$T^\varphi:\{0,1\}^*\times\mathbb N\to\mathbb Q$ is total
and belongs to $\operatorname{Name}(\mu)$.
Here, $T^\varphi$ denotes $T$ with oracle access to $\varphi$.
Define $H(\rho):=\{\mu:\mu\preceq_{\mathrm{pred}}\rho\}$.
\end{definition}

The same machine must compute a name of $\mu$ from every
name of $\rho$. Note that it is possible for different names to use different rational approximations and to represent
conditional presentations that differ at histories of zero $\rho$-probability. These conventions are borrowed from the standard treatment of real-valued
inputs in computable analysis \cite{Weihrauch2000}.
The output names may differ, but they must all represent
the same measure $\mu$. In particular, they may assign
different conditional probabilities at histories $\sigma$
with $\mu(\sigma)=0$, where these choices do not affect
the measure.

\begin{proposition}
\label{prop:forecast-closure}
Let $\rho\in\mathcal M_{\mathrm{refl}}^O$. If
$\mu\preceq_{\mathrm{pred}}\rho$, then
$\mu\in\mathcal M_{\mathrm{refl}}^O$. Consequently, $
    H(\xi^O)\subseteq\mathcal M_{\mathrm{refl}}^O$.
\end{proposition}

\begin{proof}
Let $T$ witness $\mu\preceq_{\mathrm{pred}}\rho$. Since $\rho$ has a conditional presentation that is uniformly computable with access to $O$, we can simulate $T$ using $O$ to answer its requests for rational approximations. We store each approximation and return the same value whenever the same history and precision are requested again. The answers are therefore consistent with a single name of $\rho$, so the simulation computes a name of $\mu$ almost surely.\\

We apply the sampling procedure described above to obtain a probabilistic Turing machine with access to $O$ that samples the conditionals given by this name. It halts almost surely at every history $\sigma$ and outputs $1$ with probability $\mu(1|\sigma)$ whenever $\mu(\sigma)>0$. Since the machine halts almost surely with binary output, reflective completion preserves its output probabilities. At histories with $\mu(\sigma)=0$, we have $\mu(\sigma a)=0$ for each $a\in\{0,1\}$, so the conditional probabilities assigned there do not affect the measure. Thus the completed measure is $\mu$, and hence $\mu\in\mathcal M_{\mathrm{refl}}^O$.
\end{proof}

With access to the predictions of $\xi^O$, a hypothesis can be specified by a rule that revises some predictions and retains the others.
Proposition~\ref{prop:forecast-closure} ensures that the resulting hypothesis is still dominated by $\xi^O$. The following proposition bounds its predictive advantage in terms of the length of a program specifying these revisions.

\begin{definition}
Fix a universal prefix-free oracle Turing machine $\mathbb U$. Programs are prefix-free descriptions of oracle Turing machines, decoded without access to the oracle. Given a program $p\in\fs$, a history $\sigma\in\fs$, a precision $n\in\mathbb N$, and oracle access to a function $\varphi:\fs\times\mathbb N\to\mathbb Q$, the machine outputs a rational number whenever it halts. For fixed $p$ and $\varphi$, we write $\mathbb U^\varphi(p)$ for the function $(\sigma,n)\mapsto\mathbb U^\varphi(p,\sigma,n)$. For a probability measure $\rho$ and a probability measure $\nu$, define
\[
    K_\rho(\nu)
    :=
    \min\bigl\{
        |p|:
        \mathbb U^\varphi(p)\in\operatorname{Name}(\nu)
        \text{ for every }\varphi\in\operatorname{Name}(\rho)
    \bigr\}.
\]
If no such program exists, set $K_\rho(\nu)=\infty$. Thus $K_\rho(\nu)<\infty$ exactly when $\nu\preceq_{\mathrm{pred}}\rho$.
\end{definition}

\begin{proposition}
\label{thm:forecast-relative-dominance}
Let $\rho\in\mathcal M_{\mathrm{refl}}^O$. There is a constant $C_\rho\geq0$ such that, for every $\nu\preceq_{\mathrm{pred}}\rho$ and every $\sigma\in\fs$,
\begin{equation}
    \xi^O(\sigma)
    \geq
    2^{-K_\rho(\nu)-C_\rho}\nu(\sigma).
    \label{eq:forecast-relative-dominance}
\end{equation}
Consequently, for every $\alpha\in\cp$ and $n\geq1$,
\begin{equation}
    \sum_{t\leq n}
    \log
    \frac{\nu(\alpha_t|\alpha_{<t})}
         {\xi^O(\alpha_t|\alpha_{<t})}
    \leq
    \bigl(K_\rho(\nu)+C_\rho\bigr)\ln 2.
    \label{eq:forecast-relative-log-bound}
\end{equation}
Here, logarithms are natural, $\log0=-\infty$, and the conditionals of $\nu$ at histories of zero $\nu$-probability may be given by any total conditional presentation.
\end{proposition}

\begin{proof}
Fix a program approximating, with access to $O$, a conditional
presentation of $\rho$, and let $p$ be a shortest description
of $\nu$ relative to $\rho$. As in the proof of Proposition~\ref{prop:forecast-closure}, simulate $\mathbb U^\varphi(p)$ by providing rational approximations to this presentation of $\rho$, storing each approximation so that repeated requests receive the same answer. Almost surely, these answers are consistent with a single name of $\rho$, so the simulation computes a name of $\nu$. The sampling procedure can be implemented by a probabilistic Turing
machine with access to $O$ that halts almost surely at every history
and samples the conditionals of $\nu$ wherever $\nu(\sigma)>0$.
The induced measure is $\nu$, and reflective completion leaves it unchanged.\\

The simulation and sampling procedures are fixed once $\rho$ is fixed. They therefore convert $p$ into a description of a machine whose completion is $\nu$, using only a constant number of additional bits. Thus there is a constant $d_\rho\geq0$, independent of $\nu$, such that $K(\nu)\leq K_\rho(\nu)+d_\rho$. Since the predictor $\xi^O$ dominates the mixture over the reflective hypothesis class,
\[
    \xi^O(\sigma)
    \geq
    2^{-K(\nu)}\nu(\sigma)
    \geq
    2^{-K_\rho(\nu)-d_\rho}\nu(\sigma).
\]
Taking $C_\rho=d_\rho$ proves \eqref{eq:forecast-relative-dominance}.\\

If $\nu(\alpha_{1:n})>0$, the chain rule and \eqref{eq:forecast-relative-dominance} give
\[
    \sum_{t\leq n}
    \log
    \frac{\nu(\alpha_t|\alpha_{<t})}
         {\xi^O(\alpha_t|\alpha_{<t})}
    =
    \log
    \frac{\nu(\alpha_{1:n})}
         {\xi^O(\alpha_{1:n})}
    \leq
    \bigl(K_\rho(\nu)+C_\rho\bigr)\ln 2.
\]
If $\nu(\alpha_{1:n})=0$, at least one conditional factor along the string is zero. Since $\xi^O$ has full support, the sum on the left is then $-\infty$. This proves \eqref{eq:forecast-relative-log-bound} in both cases.
\end{proof}

\subsection{Incomplete Models}

A predictive strategy may make predictions only at selected positions in a sequence. In an online classification task, for example, a predictor receives successive input--label pairs and predicts each label after observing its input, without predicting the inputs themselves. Such a predictor may correctly determine the labels by a computable rule even when the inputs are selected adversarially. It is desirable for a theory of universal induction to account for learning such rules without requiring a computable model of how the unrestricted observations are generated.\\

Lattimore, Hutter, and Gavane \cite{LattimoreHutterGavane2011} show that multiplicative dominance over computable sources does not suffice to learn computable regularities on selected subsequences. In particular, they construct a sequence $\alpha$ satisfying $\alpha_{2n}=\alpha_{2n-1}$ for every $n$ such that $\liminf_n M(\alpha_{2n}\mid\alpha_{<2n})<1$. The construction uses $M$'s own predictions to choose the unrestricted odd bits, while setting each even bit equal to its predecessor. The odd bits are chosen so that, at infinitely many even positions, $M$ assigns the correct bit a probability bounded away from $1$, despite the rule holding throughout.\\

Such partial rules specify a family of compatible probability laws without assigning prior probabilities to its members. Kosoy \cite{kosoy2017} formalizes incomplete models as convex\footnote{A set $C$ of probability measures on $\cp$ is convex if $\theta\mu+(1-\theta)\nu\in C$ for every $\mu,\nu\in C$ and every $\theta\in[0,1]$. Thus a mixture of two measures in $C$ also belongs to $C$.} sets of probability measures. A probability distribution on $\{0,1\}$ is determined by its probability of $1$, so every nonempty closed convex\footnote{A convex set $C$ of probability measures on $\cp$ is closed in the weak topology if, whenever a sequence $(\mu_n)$ in $C$ converges on every cylinder to a probability measure $\mu$, we have $\mu\in C$. Such a set is called closed convex.} set of binary distributions corresponds to an interval $[l,u]\subseteq[0,1]$. We consider bounds $l(\sigma)$ and $u(\sigma)$ on the next-bit probability at each history $\sigma$. The resulting family of compatible laws is convex because the defining inequalities $l(\sigma)\mu(\sigma)\leq\mu(\sigma1)\leq u(\sigma)\mu(\sigma)$ are linear in $\mu$. The following theorem concerns incomplete models of this form. Kosoy's main theorem \cite[Theorem~1]{kosoy2017} constructs a forecaster that converges almost surely to compatible incomplete models, for any countable family satisfying the stated conditional-model assumptions, in an appropriately normalized Kantorovich--Rubinstein metric. Our models have the more specific form of computable interval constraints. For these models, we establish convergence for the reflective Solomonoff predictor $\xi^O$ itself and bound its expected cumulative squared error in terms of a description of the bounds.
\begin{theorem}
\label{thm:incomplete-models}
Suppose $l,u:\fs\to[0,1]$ are computable and $l(\sigma)\leq u(\sigma)$ for every $\sigma$. Define
\[
I=\bigl\{\mu\in\Delta(\cp):
l(\sigma)\mu(\sigma)\leq\mu(\sigma1)\leq u(\sigma)\mu(\sigma)
\text{ for every }\sigma\in\fs\bigr\}.
\]
For $p\in[0,1]$ and $0\leq a\leq b\leq1$, write
$\operatorname{dist}(p,[a,b])=\max\{a-p,p-b,0\}$, and set
\[
d_t(\alpha)=\operatorname{dist}\!\left(
\xi^O(1|\alpha_{<t}),[l(\alpha_{<t}),u(\alpha_{<t})]\right).
\]
There is a constant $C\geq0$, depending only on the fixed predictor and coding conventions, such that for every such pair $l,u$ and every $\mu\in I$,
\[
2\mathbb E_\mu\!\left[\sum_{t=1}^{\infty}d_t(\alpha)^2\right]
\leq \bigl(K(l,u)+C\bigr)\ln 2.
\]
Here $K(l,u)$ is the length of a shortest prefix-free description of a program computing names of both bounds, uniformly in the history and requested precision. In particular, $d_t(\alpha)\to0$ $\mu$-almost surely.
\end{theorem}

\begin{proof}

Let $\nu_I(1|\sigma)=\max\{l(\sigma),\min\{\xi^O(1|\sigma),u(\sigma)\}\}$. Set $\nu_I(0|\sigma)=1-\nu_I(1|\sigma)$ and extend by the chain rule. The projection is computable from names of $l,u$ and $\xi^O$, so $\nu_I\preceq_{\mathrm{pred}}\xi^O$ and Proposition~\ref{prop:forecast-closure} gives $\nu_I\in\mathcal M^O_{\mathrm{refl}}$. A fixed program implements this projection once a description of $l,u$ is given. Hence, for a constant $c$ independent of $l,u$,
\[
K_{\xi^O}(\nu_I)\leq K(l,u)+c.
\]
Set $B_I=(K_{\xi^O}(\nu_I)+C_{\xi^O})\ln 2$, where $C_{\xi^O}$ is the constant in Proposition~\ref{thm:forecast-relative-dominance}. That proposition gives, for every $\alpha\in\cp$ and $n\geq1$,
\[
\sum_{t\leq n}
\log\frac{\nu_I(\alpha_t|\alpha_{<t})}{\xi^O(\alpha_t|\alpha_{<t})}
\leq B_I\leq\bigl(K(l,u)+C\bigr)\ln 2,
\]
where $C=c+C_{\xi^O}$ and $\log0=-\infty$.

Fix $\mu\in I$. For every history $\sigma$ with $\mu(\sigma)>0$, we have $\mu(1|\sigma)\in[l(\sigma),u(\sigma)]$. The definition of $\nu_I$ also ensures that $\nu_I(x|\sigma)=0$ implies $\mu(x|\sigma)=0$ at such histories. Consequently, $\nu_I(\sigma)>0$ whenever $\mu(\sigma)>0$. The logarithmic terms are, therefore, finite $\mu$-almost surely and have finite expectations for each fixed $n$. In the sums below, terms with $\mu(x|\alpha_{<t})=0$ are taken to be zero. Taking expectations with respect to $\mu$ and conditioning on the history, we obtain
\begin{equation}
\label{eq:incomplete-log-bound}
\begin{aligned}
\sum_{t\leq n}
\mathbb{E}_{\mu}\!\left[
\log \frac{\nu_I(\alpha_t|\alpha_{<t})}
{\xi^O(\alpha_t|\alpha_{<t})}
\right]
&=
\sum_{t\leq n}
\mathbb{E}_{\mu}\!\left[
\mathbb{E}_{\mu}\!\left[
\left.
\log \frac{\nu_I(\alpha_t|\alpha_{<t})}
{\xi^O(\alpha_t|\alpha_{<t})}
\right|
\alpha_{<t}
\right]
\right] \\
&=
\sum_{t\leq n}
\mathbb{E}_{\mu}\!\left[
\sum_{x\in\{0,1\}}
\mu(x|\alpha_{<t})
\log
\frac{\nu_I(x|\alpha_{<t})}
{\xi^O(x|\alpha_{<t})}
\right] \\
&=
\sum_{t\leq n}
\mathbb{E}_{\mu}\!\left[
D_{\mathrm{KL}}\!\left(
\mu(\cdot|\alpha_{<t})
\middle\|
\xi^O(\cdot|\alpha_{<t})
\right)
-
D_{\mathrm{KL}}\!\left(
\mu(\cdot|\alpha_{<t})
\middle\|
\nu_I(\cdot|\alpha_{<t})
\right)
\right] \\
&\leq B_I.
\end{aligned}
\end{equation}

We next compare the difference of divergences with $D_{\mathrm{KL}}(\nu_I(\cdot|\sigma)\|\xi^O(\cdot|\sigma))$ at histories with $\mu(\sigma)>0$. If $\nu_I(1|\sigma)=0$, then $u(\sigma)=0$, so $\mu(1|\sigma)=0$. Similarly, if $\nu_I(1|\sigma)=1$, then $l(\sigma)=1$, so $\mu(1|\sigma)=1$. In either case, $\mu(\cdot|\sigma)=\nu_I(\cdot|\sigma)$, and the required inequality holds with equality.

For the remaining histories, $0<\nu_I(1|\sigma)<1$. Since $\mu(\cdot|\sigma)$, $\xi^O(\cdot|\sigma)$, and $\nu_I(\cdot|\sigma)$ are Bernoulli measures on $\{0,1\}$, the following identity applies:
\begin{equation}
\label{eq:incomplete-kl-identity}
\begin{gathered}
D_{\mathrm{KL}}\!\left(
\mu(\cdot|\alpha_{<t})
\middle\|
\xi^O(\cdot|\alpha_{<t})
\right)
-
D_{\mathrm{KL}}\!\left(
\mu(\cdot|\alpha_{<t})
\middle\|
\nu_I(\cdot|\alpha_{<t})
\right)
-
D_{\mathrm{KL}}\!\left(
\nu_I(\cdot|\alpha_{<t})
\middle\|
\xi^O(\cdot|\alpha_{<t})
\right)
\\[4pt]
=
\left(
\mu(1|\alpha_{<t})
-
\nu_I(1|\alpha_{<t})
\right)
\log\!\left(
\frac{
\nu_I(1|\alpha_{<t})
\left(1-\xi^O(1|\alpha_{<t})\right)
}{
\xi^O(1|\alpha_{<t})
\left(1-\nu_I(1|\alpha_{<t})\right)
}
\right).
\end{gathered}
\end{equation}

The right-hand side of \eqref{eq:incomplete-kl-identity} is non-negative. There are three cases to consider:
\begin{enumerate}
    \item $\xi^O(1|\alpha_{<t})\in[l(\alpha_{<t}),u(\alpha_{<t})]$, implying that $\nu_I(1|\alpha_{<t})=\xi^O(1|\alpha_{<t})$. Here, the logarithmic factor is zero.

    \item $\xi^O(1|\alpha_{<t})<l(\alpha_{<t})$, implying that $\nu_I(1|\alpha_{<t})=l(\alpha_{<t})$. Then $\nu_I(1|\alpha_{<t})-\xi^O(1|\alpha_{<t})>0$, so the logarithmic factor is positive. Since $\mu\in I$ gives $\mu(1|\alpha_{<t})\geq l(\alpha_{<t})=\nu_I(1|\alpha_{<t})$, the factor $\mu(1|\alpha_{<t})-\nu_I(1|\alpha_{<t})$ is non-negative. Therefore, the product is non-negative.

    \item $\xi^O(1|\alpha_{<t})>u(\alpha_{<t})$, implying that $\nu_I(1|\alpha_{<t})=u(\alpha_{<t})$. Then $\nu_I(1|\alpha_{<t})-\xi^O(1|\alpha_{<t})<0$, so the logarithmic factor is negative. Since $\mu\in I$ gives $\mu(1|\alpha_{<t})\leq u(\alpha_{<t})=\nu_I(1|\alpha_{<t})$, the factor $\mu(1|\alpha_{<t})-\nu_I(1|\alpha_{<t})$ is non-positive. Therefore, the product is non-negative.
\end{enumerate}

Together with the endpoint cases, this yields
\[
D_{\mathrm{KL}}\!\left(
\mu(\cdot|\alpha_{<t})
\middle\|
\xi^O(\cdot|\alpha_{<t})
\right)
-
D_{\mathrm{KL}}\!\left(
\mu(\cdot|\alpha_{<t})
\middle\|
\nu_I(\cdot|\alpha_{<t})
\right)
\geq
D_{\mathrm{KL}}\!\left(
\nu_I(\cdot|\alpha_{<t})
\middle\|
\xi^O(\cdot|\alpha_{<t})
\right).
\]

Combining this inequality with \eqref{eq:incomplete-log-bound}, we obtain
\[
\sum_{t\leq n}
\mathbb{E}_{\mu}\!\left[
D_{\mathrm{KL}}\!\left(
\nu_I(\cdot|\alpha_{<t})
\middle\|
\xi^O(\cdot|\alpha_{<t})
\right)
\right]
\leq B_I.
\]

By Pinsker's inequality, for each $t$,
\[
D_{\mathrm{KL}}\!\left(
\nu_I(\cdot|\alpha_{<t})
\middle\|
\xi^O(\cdot|\alpha_{<t})
\right)
\geq
2\left|
\nu_I(1|\alpha_{<t})
-
\xi^O(1|\alpha_{<t})
\right|^2.
\]
Therefore,
\[
2\sum_{t\leq n}
\mathbb{E}_{\mu}\!\left[
\left|
\nu_I(1|\alpha_{<t})
-
\xi^O(1|\alpha_{<t})
\right|^2
\right]
\leq B_I.
\]

By the definition of $\nu_I$,
\[
\left|
\nu_I(1|\alpha_{<t})
-
\xi^O(1|\alpha_{<t})
\right|
=
\max\left\{
l(\alpha_{<t})-\xi^O(1|\alpha_{<t}),
\,
\xi^O(1|\alpha_{<t})-u(\alpha_{<t}),
\,
0
\right\}.
\]
It follows that
\[
2\sum_{t\leq n}
\mathbb{E}_{\mu}\!\left[
\left(
\max\left\{
l(\alpha_{<t})-\xi^O(1|\alpha_{<t}),
\,
\xi^O(1|\alpha_{<t})-u(\alpha_{<t}),
\,
0
\right\}
\right)^2
\right]
\leq B_I.
\]
Letting $n\to\infty$ and applying the monotone convergence theorem produces:
\[
2\mathbb{E}_{\mu}\!\left[
\sum_{t=1}^{\infty}
\left(
\max\left\{
l(\alpha_{<t})-\xi^O(1|\alpha_{<t}),
\,
\xi^O(1|\alpha_{<t})-u(\alpha_{<t}),
\,
0
\right\}
\right)^2
\right]
\leq B_I.
\]

Since this non-negative series has finite expectation, it is finite $\mu$-almost surely. We conclude that $\max\{l(\alpha_{<t})-\xi^O(1|\alpha_{<t}),\,\xi^O(1|\alpha_{<t})-u(\alpha_{<t}),\,0\}\to0$ as $t\to\infty$, $\mu$-almost surely. In other words, the distance between the forecast and the interval prescribed at each history converges to zero.

Moreover, for every $\varepsilon>0$, a term exceeds $\varepsilon$ only when its square exceeds $\varepsilon^2$, so
\[
\mathbb{E}_{\mu}\!\left[
\left|
\left\{
t\in\mathbb{N}:
\operatorname{dist}\!\left(
\xi^O(1|\alpha_{<t}),
[l(\alpha_{<t}),u(\alpha_{<t})]
\right)>\varepsilon
\right\}
\right|
\right]
\leq
\frac{B_I}{2\varepsilon^2},
\]
bounding the expected number of $\varepsilon$-violations.

\end{proof}

Theorem~\ref{thm:finite-horizon-incomplete-models} in the appendix extends this result to computably specified events concerning the next $m$ bits, for any fixed $m\geq1$. Lastly, the convergence conclusion also holds when $l$ and $u$ are computable with access to $O$, since the sampling property still ensures that $\nu_I\in\mathcal M_{\mathrm{refl}}^O$, although allowing the model's restrictions to depend on the oracle seems rather unnatural. 

\subsection{Prediction of Selected Bits}
\label{subsec:selected-bits}

The following corollaries establish the success of reflective Solomonoff induction on selected bits. They extend the analysis of Lattimore, Hutter, and Gavane \cite{LattimoreHutterGavane2011} to the reflective setting, with guarantees for both deterministic and stochastic rules.

\begin{corollary}
\label{cor:deterministic-selected-bits}
Let $f:\{0,1\}^{*}\to\{0,1,\epsilon\}$ be a total computable function, where $\epsilon$ indicates that no prediction is made, and let $\alpha\in\{0,1\}^{\infty}$ satisfy $f$ in the sense that $f(\alpha_{<n})=\alpha_n$ whenever $f(\alpha_{<n})\neq\epsilon$. Suppose $f$ makes predictions at infinitely many positions along $\alpha$, and let $p_1<p_2<\cdots$ enumerate these positions. Then $\xi^O$ predicts correctly along this subsequence:
\[
\lim_{i\to\infty}\xi^O(\alpha_{p_i}|\alpha_{<p_i})=1.
\]
\end{corollary}

\begin{proof}
    We set $l(\sigma)=u(\sigma)=f(\sigma)$ whenever $f(\sigma)\neq\epsilon$, and $l(\sigma)=0$ and $u(\sigma)=1$ otherwise. In line with the statement of Theorem \ref{thm:incomplete-models}, the bounds are computable and define an incomplete model $I$.

    Consider the probability measure $\delta_{\alpha}$ that assigns all its probability mass to $\alpha$. At each prefix $\sigma=\alpha_{<t}$, we have
$\delta_\alpha(1\mid\sigma)=\alpha_t$. Since $\alpha$ satisfies $f$,
this agrees with $f(\sigma)$ whenever $f(\sigma)\neq\epsilon$.
At every other history $\sigma$, we have
$\delta_\alpha(\sigma)=\delta_\alpha(\sigma1)=0$. In other words, $\delta_\alpha\in I$. Theorem \ref{thm:incomplete-models} ensures that $d_t(\beta)\to0$ for $\delta_\alpha$-almost
every $\beta$. But since every set of $\delta_\alpha$-measure one contains
$\alpha$, we have $d_t(\alpha)\to0$. At each selected bit $p_i$, the prescribed interval is
$\{\alpha_{p_i}\}$, so $d_{p_i}(\alpha)
=
\left|\xi^O(1\mid\alpha_{<p_i})-\alpha_{p_i}\right|
=
1-\xi^O(\alpha_{p_i}\mid\alpha_{<p_i})$.
Since $p_i\to\infty$, it follows that
$\xi^O(\alpha_{p_i}\mid\alpha_{<p_i})\to1$.
\end{proof}

Lattimore, Hutter, and Gavane \cite[Section~5]{LattimoreHutterGavane2011} ask whether their deterministic selected-bits theorem for normalized ordinary Solomonoff induction extends to computable stochastic patterns. The following corollary answers the corresponding question for reflective Solomonoff induction; it does not resolve the question for the ordinary predictor.

\begin{corollary}
\label{cor:stochastic-selected-bits}
Let $g:\{0,1\}^{*}\to\Delta(\{0,1\})\cup\{\epsilon\}$ be a computable rule: the set $\{\sigma:g(\sigma)\neq\epsilon\}$ is decidable and, on this set, the probabilities $g(\sigma)(a)$ are uniformly computable in $\sigma$ and $a$. Let $\mu$ be a probability measure on $\{0,1\}^{\infty}$ such that $\mu(a|\sigma)=g(\sigma)(a)$ for each $a\in\{0,1\}$ whenever $\mu(\sigma)>0$ and $g(\sigma)\neq\epsilon$. Then, for $\mu$-almost every $\alpha$ along which $g$ makes infinitely many predictions, enumerated at positions $p_1<p_2<\cdots$, we have
\[
\lim_{i\to\infty}
\bigl(\xi^O(\alpha_{p_i}|\alpha_{<p_i})
-g(\alpha_{<p_i})(\alpha_{p_i})\bigr)=0.
\]
\end{corollary}

\begin{proof}
Set $l(\sigma)=u(\sigma)=g(\sigma)(1)$ whenever $g(\sigma)\neq\epsilon$, and set $l(\sigma)=0$ and $u(\sigma)=1$ otherwise. These bounds are computable by the assumptions on $g$. The assumption on $\mu$ places it in the corresponding incomplete model.\\

By Theorem~\ref{thm:incomplete-models}, the distance from $\xi^O(1|\alpha_{<n})$ to $[l(\alpha_{<n}),u(\alpha_{<n})]$ converges to zero $\mu$-almost surely. At each selected position $p_i$, this distance equals $|\xi^O(1|\alpha_{<p_i})-g(\alpha_{<p_i})(1)|$. Since the probabilities of $0$ are the complements of those of $1$, this also equals $|\xi^O(\alpha_{p_i}|\alpha_{<p_i})-g(\alpha_{<p_i})(\alpha_{p_i})|$. The conclusion follows.
\end{proof}

\subsection{Unbiasedness and calibration}

Calibration connects probability assignments with observed frequencies. Among events repeatedly assigned probability $p$, the proportion that occur should approach $p$. Dawid \cite{Dawid1982} shows that a Bayesian predictor is almost surely calibrated under its own probability law. We establish a weighted form of calibration for $\xi^O$ that holds on every sequence.\\

Calibration through continuous tests is central to Kakade and Foster's deterministic weak calibration \cite{KakadeFoster2004} and to defensive forecasting \cite{VovkTakemuraShafer2005}, where forecasts are chosen to prevent continuous betting strategies from accumulating unbounded capital. Garrabrant et al.\ \cite[Section~4.3]{GarrabrantEtAl2016} establish related weighted-unbiasedness and calibration properties for logical inductors. Their recurring-unbiasedness result gives zero as a limit point, while their unbiasedness-from-feedback theorem gives convergence under a feedback condition. In our sequential setting each outcome is observed before the next prediction. The following argument proves convergence for $\xi^O$ by representing multiplicative betting strategies as comparison measures and applying the dominance bound.\\

We first consider unbiasedness, understood here as convergence of the average difference between outcomes and predictions to zero. This permits overprediction in one probability range to offset underprediction in another. To distinguish these errors, we consider weighted averages, with weights depending on the predicted probabilities. We show that the weighted average error of $\xi^O$ converges to zero for each admissible weighting scheme whenever its accumulated weight diverges. Choosing continuous weights concentrated on a probability interval then yields the corresponding calibration property.

\begin{definition}[Admissible weighting scheme]
\label{def:admissible-weighting}
A function $w:[0,1]\to[0,1]$ is an admissible weighting scheme if it is computable as a real-valued function: a single procedure computes a name of $w(p)$ from every Cauchy name of $p$, uniformly in the requested precision. 
\end{definition}

\begin{proposition}[Unbiasedness]
\label{unbiased}
Let $\alpha\in\cp$ and let $w$ be an admissible weighting scheme. Set
$p_t=\xi^O(1|\alpha_{<t})$, $w_t=w(p_t)$, and $W_n=\sum_{t\leq n}w_t$.
If $W_n\to\infty$, then
\[
\lim_{n\to\infty}\frac{1}{W_n}\sum_{t\leq n}w_t(\alpha_t-p_t)=0.
\]
\end{proposition}

\begin{proof}
Fix a rational $y\in[-1/2,1/2]$. At each history $\sigma$, write $p_\sigma=\xi^O(1|\sigma)$ and define
\[
\mu_y(x|\sigma)=\xi^O(x|\sigma)
\bigl(1+yw(p_\sigma)(x-p_\sigma)\bigr),\qquad x\in\{0,1\}.
\]
The factor in parentheses lies in $[1/2,3/2]$, so
$\mu_y(x\mid\sigma)\geq0$ for each $x\in\{0,1\}$. Moreover,
\[
\sum_{x\in\{0,1\}}\mu_y(x|\sigma)
=1+yw(p_\sigma)\bigl(p_\sigma(1-p_\sigma)-(1-p_\sigma)p_\sigma\bigr)=1.
\]
Thus these conditionals define a probability measure. Since $w$ is computable, $\mu_y\preceq_{\mathrm{pred}}\xi^O$. Proposition~\ref{thm:forecast-relative-dominance} gives
\[
\sum_{t\leq n}\log\bigl(1+yw_t(\alpha_t-p_t)\bigr)
\leq A_y,
\qquad
A_y=\bigl(K_{\xi^O}(\mu_y)+C_{\xi^O}\bigr)\ln 2.
\]
Let $S_n=\sum_{t\leq n}w_t(\alpha_t-p_t)$. The inequality
$\log(1+v)\geq v-v^2$ for $|v|\leq1/2$ yields
\[
yS_n\leq A_y+y^2\sum_{t\leq n}w_t^2(\alpha_t-p_t)^2
\leq A_y+y^2W_n,
\]
since $0\leq w_t\leq1$ and $|\alpha_t-p_t|\leq1$.
For every rational $y\in(0,1/2]$, applying this bound to $y$ and $-y$ gives, whenever $W_n>0$,
\[
-\frac{A_{-y}}{yW_n}-y
\leq\frac{S_n}{W_n}
\leq\frac{A_y}{yW_n}+y.
\]
As $W_n\to\infty$, it follows that
$-y\leq\liminf_n S_n/W_n\leq\limsup_n S_n/W_n\leq y$.
Taking positive rational $y$ arbitrarily close to zero proves the claim.
\end{proof}

\begin{definition}
Following Garrabrant et al.\ \cite[Definition~4.3.2]{GarrabrantEtAl2016}, let $\delta>0$ be rational and let $x,y\in\mathbb R$. Define
\[
\operatorname{Ind}_\delta(x>y)
:=
\begin{cases}
0,&x\leq y,\\[2mm]
\dfrac{x-y}{\delta},&y<x\leq y+\delta,\\[3mm]
1,&y+\delta<x.
\end{cases}
\]
\end{definition}

\begin{proposition}[Calibration]
\label{claim:calibration}
Let $\alpha\in\cp$, let $a<b$ be computable reals in $[0,1]$, and let $\delta>0$ be rational. Define
\[
w(p)=\min\{\operatorname{Ind}_\delta(p>a),\operatorname{Ind}_\delta(b>p)\},
\qquad p\in[0,1].
\]
Set $p_t=\xi^O(1|\alpha_{<t})$, $w_t=w(p_t)$, and $W_n=\sum_{t\leq n}w_t$.
If $W_n\to\infty$, then
\[
\frac{\sum_{t\leq n}w_t\alpha_t}{W_n}
-\frac{\sum_{t\leq n}w_tp_t}{W_n}\longrightarrow0.
\]
Every limit point of $\bigl(\sum_{t\leq n}w_t\alpha_t\bigr)/W_n$ therefore belongs to $[a,b]$. If either weighted average converges, both converge to the same value.
\end{proposition}

\begin{proof}
The function $w$ is computable and takes values in $[0,1]$, so it is an admissible weighting scheme. Proposition~\ref{unbiased} gives the asserted convergence of the difference. Whenever $w_t>0$, we have $a<p_t<b$. Hence, for all $n$ with $W_n>0$,
\[
a\leq\frac{\sum_{t\leq n}w_tp_t}{W_n}\leq b.
\]
The distance of the weighted empirical frequency from $[a,b]$ is at most the absolute difference of the two averages, which tends to zero. This proves the claim about limit points. The equality of limits when either average converges follows from the same difference estimate.
\end{proof}

The weighted empirical frequency $\sum_{t\leq n}w_t\alpha_t/W_n$
is the total weight assigned to observations of $1$, divided by
the total weight assigned to all observations. The weights restrict
attention to predictions in $[a,b]$, giving less weight to predictions
near the endpoints. The proposition says that, whenever the accumulated weight diverges,
the weighted empirical frequency of $1$s among observations whose
predicted probability of $1$ lies in $[a,b]$ eventually lies arbitrarily
close to the interval $[a,b]$.

\section{A predictor presenting its hypothesis class}
\label{subsec:class-presentation}

The preceding results establish $ H(\xi^O)\subseteq\mathcal M_{\mathrm{refl}}^O$. This inclusion was sufficient for the predictive applications: hypotheses constructed using the predictor's forecasts remain within its dominated class. We now ask whether the converse can also hold. Can every reflective hypothesis be described using the predictor's forecasts? We say that a predictor $\rho$ presents a hypothesis class $\mathcal C$ when $H(\rho)=\mathcal C$. Such a presentation makes the predictor's probabilities sufficient for describing every hypothesis in the class. For arbitrary measures, multiplicative dominance does not imply this access: two measures may assign comparable probabilities to every finite string while carrying different computational information.

\begin{proposition}
\label{prop:dominance-recoverability}
For every full-support probability measure $\rho$, there is a full-support probability measure $\mu$ such that $
    \frac12\rho(\sigma)
    \leq
    \mu(\sigma)
    \leq
    2\rho(\sigma)$ for every $\sigma\in\fs$ but $\mu\not\preceq_{\mathrm{pred}}\rho$.
\end{proposition}

\begin{proof}
Fix a name $\varphi\in\operatorname{Name}(\rho)$ and choose $u\in\cp$ that is not computable from $\varphi$. We will define $\mu$ so that each bit $u_n$ can be recovered by comparing the conditional probabilities of $\mu$ and $\rho$ at the history $r_n=1^{n-1}0$. The set $R=\{r_n:n\geq1\}$ is prefix-free, so a path passes through at most one of these histories.\\

Let $\lambda_0=1/2$ and $\lambda_1=2$. Define $\mu(1|\sigma)=\rho(1|\sigma)$ whenever $\sigma\notin R$, and set 
\[\mu(1| r_n)
    =
    \frac{\lambda_{u_n}\rho(1| r_n)}
         {1-\rho(1| r_n)+\lambda_{u_n}\rho(1| r_n)}\]
For every history $\sigma$, set $\mu(0|\sigma)=1-\mu(1|\sigma)$. Define $\mu(\epsilon)=1$ and recursively set $\mu(\sigma a)=\mu(\sigma)\mu(a|\sigma)$ for $a\in\{0,1\}$. At each modified history $r_n$, the assumption $0<\rho(1| r_n)<1$ ensures that conditional probabilities are strictly positive. Elsewhere, they agree with those of $\rho$. Consequently, $\mu(\sigma)>0$ for every finite string $\sigma$, so $\mu$ has full support. At each modified history, the definition of $\mu$ ensures that the ratio $\dfrac{\mu(\cdot|r_n)}{\rho(\cdot|r_n)}$ lies in $[1/2,2]$ for either next bit. All other conditional ratios equal $1$. By the chain rule, $\mu(\sigma)/\rho(\sigma)$ is the product of the conditional ratios along $\sigma$. Since $R$ is prefix-free, at most one prefix of $\sigma$ belongs to $R$, so at most one factor in this product differs from $1$. The product therefore also lies in $[1/2,2]$, giving $\frac12\rho(\sigma)\leq\mu(\sigma)\leq2\rho(\sigma)$.\\

Suppose now that $\mu\preceq_{\mathrm{pred}}\rho$. Applying a witnessing procedure to $\varphi$ gives a name for $\mu$. From this name and $\varphi$, we can compute $
    \dfrac{\mu(1| r_n)\rho(0| r_n)}
         {\mu(0| r_n)\rho(1| r_n)}
    =
    \lambda_{u_n}$. The denominators are positive, and the two possible values, $1/2$ and $2$, are separated by a fixed gap. We can therefore determine each $u_n$ from sufficiently accurate approximations. This computes $u$ from $\varphi$. We obtain a contradiction.
\end{proof}

Proposition~\ref{prop:dominance-recoverability} concerns arbitrary measures. Its construction does not ensure that $\mu$ belongs to $\mathcal M^O_{\mathrm{refl}}$ when $\rho$ does. It therefore does not establish a separation between dominance and recoverability within the reflective class. Whether the fixed predictor $\xi^O$ presents the class, and whether some dominant member of the class fails to do so, remain unresolved here.\\

We now show that a dominant predictor can be chosen to present the reflective class. The construction records the conditional probabilities of each reflective hypothesis at designated histories. These histories are chosen so that at most one prediction changes along any sequence, while maintaining optimality.

\begin{theorem}
\label{thm:reflective-class-presentation}
There is a full-support measure
$\widehat{\xi}^O\in\mathcal M_{\mathrm{refl}}^O$ such that $H(\widehat{\xi}^O)
    =
    \mathcal M_{\mathrm{refl}}^O$ and
$\widehat{\xi}^O(\sigma)
    \geq
    \frac13\xi^O(\sigma)$ for every $\sigma\in\fs$.
In particular, $\widehat{\xi}^O$ and $\xi^O$ mutually multiplicatively dominate one another.
\end{theorem}

\begin{proof}
Let $(\bar{\lambda}_e^O)_{e\in\mathbb N}$ be the enumeration of completed measures in $\mathcal M_{\mathrm{refl}}^O$, and let $p_e$ be the total conditional presentation corresponding to the completion of the corresponding oracle program, including at histories of zero $\bar{\lambda}_e^O$-probability. The values $p_e(\sigma)$ are uniformly computable with access to $O$. Fix a computable bijection $\langle\cdot,\cdot\rangle:\mathbb N\times\fs\to\mathbb N_{\geq1}$, and again let $r_n=1^{n-1}0$ for $n\geq1$ and $R=\{r_n:n\geq1\}$. The set $R$ is decidable and prefix-free. We use the history $r_{\langle e,\sigma\rangle}$ to record $p_e(\sigma)$.\\

Now, define $\widehat{\xi}^O(1| r_{\langle e,\sigma\rangle})=\frac13+\frac13p_e(\sigma)$ and $\widehat{\xi}^O(1| y)=\xi^O(1| y)$ whenever $y\notin R$. For every history $\sigma$, set $\widehat{\xi}^O(0|\sigma)=1-\widehat{\xi}^O(1|\sigma)$. Define $\widehat{\xi}^O(\epsilon)=1$ and recursively set $\widehat{\xi}^O(\sigma a)=\widehat{\xi}^O(\sigma)\widehat{\xi}^O(a|\sigma)$ for $a\in\{0,1\}$. Since the conditional probabilities sum to one at every history, these assignments determine a probability measure. These conditionals are uniformly computable with access to $O$. The sampling property, therefore, gives an oracle machine that halts almost surely at every history and has exactly these output probabilities. Its completion induces $\widehat{\xi}^O$, so $\widehat{\xi}^O\in\mathcal M_{\mathrm{refl}}^O$. Both conditional probabilities at a coding history lie in $[1/3,2/3]$; elsewhere they agree with those of the full-support measure $\xi^O$. Consequently, $\widehat{\xi}^O(\sigma)>0$ for every finite string $\sigma$, so $\widehat{\xi}^O$ has full support.\\

For every $r_n\in R$ and $a\in\{0,1\}$, we have $\widehat{\xi}^O(a| r_n)\geq\frac13\geq\frac13\xi^O(a| r_n)$. At all other histories, the conditional probabilities agree. By the chain rule, $\widehat{\xi}^O(\sigma)/\xi^O(\sigma)$ is the product of the conditional ratios along $\sigma$. Since $R$ is prefix-free, at most one factor differs from $1$, and that factor is at least $1/3$. Hence $\widehat{\xi}^O(\sigma)\geq\frac13\xi^O(\sigma)$ for every $\sigma\in\fs$. Thus $\widehat{\xi}^O$ dominates every member of the reflective class. Conversely, $\xi^O$ dominates $\widehat{\xi}^O$ because $\widehat{\xi}^O$ belongs to that class.\\

It remains to prove the representation claim. By the definition of $\widehat{\xi}^O$,
\[
p_e(\sigma)=3\widehat{\xi}^O(1|r_{\langle e,\sigma\rangle})-1.
\]
To approximate $p_e(\sigma)$ within $2^{-k}$, it is enough to approximate the conditional probability on the right within $2^{-(k+2)}$, since $3\cdot2^{-(k+2)}\leq2^{-k}$. This procedure works with every name for $\widehat{\xi}^O$ and uniformly in $e$. It therefore computes a name for each $\bar{\lambda}_e^O$, proving $\mathcal M_{\mathrm{refl}}^O\subseteq H(\widehat{\xi}^O)$. The reverse inclusion follows from Proposition~\ref{prop:forecast-closure}, since $\widehat{\xi}^O$ has full support and belongs to $\mathcal M_{\mathrm{refl}}^O$.
\end{proof}

The construction changes at most one next-bit prediction along any infinite sequence, although recovering a hypothesis may require querying predictions at histories that are never observed. Every $\nu\in\mathcal M_{\mathrm{refl}}^O$, therefore, has a finite description using the predictions of $\widehat{\xi}^O$. By Proposition~\ref{thm:forecast-relative-dominance} and the dominance bounds, $\widehat{\xi}^O(\sigma)\geq\widehat c_O2^{-K_{\widehat{\xi}^O}(\nu)}\nu(\sigma)$ for some constant $\widehat c_O>0$ independent of $\nu$ and $\sigma$.\\

For this predictor, the two description lengths agree up to an additive constant:
\[
K_{\widehat{\xi}^O}(\nu)=K(\nu)+O(1),\qquad \nu\in\mathcal M^O_{\mathrm{refl}}.
\]
Here $K(\nu)=\min\{K(e):\bar\lambda_e^O=\nu\}$ refers to the fixed enumeration of completed oracle programs. Given a shortest description of such an index $e$, the decoding procedure in Theorem~\ref{thm:reflective-class-presentation} computes a name of $\nu$ from any name of $\widehat{\xi}^O$, with fixed additional program length. Thus $K_{\widehat{\xi}^O}(\nu)\leq K(\nu)+O(1)$. Conversely, the compilation argument in the proof of Proposition~\ref{thm:forecast-relative-dominance}, applied with $\rho=\widehat{\xi}^O$, gives $K(\nu)\leq K_{\widehat{\xi}^O}(\nu)+O(1)$. The constants are independent of $\nu$, with the oracle, predictor, and coding conventions fixed. The predictor-relative bound therefore expresses the same complexity scale using descriptions from the predictor's probabilities.\\

Although the construction of $\widehat{\xi}^O$ depends on the chosen encoding, the following proposition shows that predictor-relative description lengths agree up to an additive constant whenever the predictors present and dominate the same class.

\begin{proposition}
\label{prop:presentation-invariance}
Suppose $\rho$ and $\eta$ have full support, satisfy $ H(\rho)=H(\eta)=\mathcal C$
and each multiplicatively dominates every member of $\mathcal C$. Then,
they mutually multiplicatively dominate one another, and there is a constant $b_{\rho,\eta}$ such that for every $\mu\in\mathcal C$, we have
\[
    |K_\rho(\mu)-K_\eta(\mu)|
    \leq
    b_{\rho,\eta}.
\]

\end{proposition}

\begin{proof}
The identity procedure gives $\rho\in H(\rho)$ and $\eta\in H(\eta)$. Since $H(\rho)=H(\eta)=\mathcal C$, each predictor is computable from the other, and both belong to $\mathcal C$. The dominance assumptions therefore imply that they mutually multiplicatively dominate one another.\\

Fix a procedure that computes a name for $\rho$ from any name for $\eta$. Composing this procedure with a shortest description of $\mu$ relative to $\rho$ gives a description of $\mu$ relative to $\eta$, with a fixed additional cost independent of $\mu$. Thus $K_\eta(\mu)\leq K_\rho(\mu)+c_{\rho,\eta}$. Reversing the roles of $\rho$ and $\eta$ gives $K_\rho(\mu)\leq K_\eta(\mu)+c_{\eta,\rho}$. Taking $b_{\rho,\eta}=\max\{c_{\rho,\eta},c_{\eta,\rho}\}$ proves the claim.
\end{proof}

The predictive results of the preceding sections rely on closure: hypotheses described using the predictor's forecasts remain available for inductive comparison. The representation theorem complements these results by providing an optimal predictor from whose probabilities every reflective hypothesis can be described. For this predictor, descriptions from its probabilities and descriptions by completed oracle programs have the same complexity up to an additive constant. Proposition~\ref{prop:presentation-invariance} gives the corresponding invariance among dominant predictors presenting the same class.

\section{Appendix}

\begin{theorem}[Finite-horizon prediction]
\label{thm:finite-horizon-incomplete-models}
Fix a horizon $m\geq1$. For each history $\sigma\in\{0,1\}^*$, let $A_\sigma\subseteq\{0,1\}^m$ specify an event concerning the next $m$ bits, with membership $\tau\in A_\sigma$ decidable uniformly in $\sigma$ and $\tau$. For any probability measure $\rho$, write $\rho(A_\sigma|\sigma)=\sum_{\tau\in A_\sigma}\rho(\sigma\tau)/\rho(\sigma)$ whenever $\rho(\sigma)>0$.

Suppose $l$ and $u$ are computable functions from $\{0,1\}^*$ to $[0,1]$ such that $l(\sigma)\leq u(\sigma)$ for every $\sigma$, and define the associated incomplete model by
\[
I=\bigl\{\mu\in\Delta(\{0,1\}^{\infty}) :
l(\sigma)\mu(\sigma)
\leq
\sum_{\tau\in A_\sigma}\mu(\sigma\tau)
\leq
u(\sigma)\mu(\sigma)
\text{ for every }\sigma\in\{0,1\}^*
\bigr\}.
\]
For $p\in[0,1]$ and $0\leq a\leq b\leq1$, let $\operatorname{dist}(p,[a,b])=\max\{a-p,p-b,0\}$.

Set
\[
d_t^{(m)}(\alpha)=\operatorname{dist}\!\left(
\xi^O(A_{\alpha_{<t}}|\alpha_{<t}),[l(\alpha_{<t}),u(\alpha_{<t})]\right).
\]
There is a constant $C\geq0$, depending only on the fixed predictor and coding conventions, such that for every such $m,A,l,u$ and every $\mu\in I$,
\[
2\mathbb E_\mu\!\left[\sum_{t=1}^{\infty}d_t^{(m)}(\alpha)^2\right]
\leq m\bigl(K(m,A,l,u)+\lceil\log_2 m\rceil+C\bigr)\ln 2.
\]
Here $A=(A_\sigma)_{\sigma\in\fs}$, and $K(m,A,l,u)$ is the length of a shortest joint prefix-free description specifying $m$, deciding membership in $A_\sigma$ uniformly in $\sigma$ and the length-$m$ string, and computing names of $l,u$. In particular, $d_t^{(m)}(\alpha)\to0$ $\mu$-almost surely.
\end{theorem}

\begin{proof}
For any probability measure $\rho$ and finite strings $\sigma$ and $\tau$ with $\rho(\sigma)>0$, write $\rho(\tau|\sigma)=\rho(\sigma\tau)/\rho(\sigma)$. Let $p_\sigma=\xi^O(A_\sigma|\sigma)$ and $q_\sigma=\max\{l(\sigma),\min\{p_\sigma,u(\sigma)\}\}$. Whenever $0<p_\sigma<1$, define a probability distribution on the next $m$ bits by
\[
b_\sigma(\tau)
=
\begin{cases}
\dfrac{q_\sigma}{p_\sigma}\xi^O(\tau|\sigma),
& \tau\in A_\sigma,\\[6pt]
\dfrac{1-q_\sigma}{1-p_\sigma}\xi^O(\tau|\sigma),
& \tau\notin A_\sigma.
\end{cases}
\]
Note that $\sum_{\tau\in A_\sigma}b_\sigma(\tau)=q_\sigma$. If $p_\sigma\in\{0,1\}$, set $b_\sigma(\tau)=\xi^O(\tau|\sigma)$. Since $\xi^O$ has full support, these cases correspond to $A_\sigma=\emptyset$ and $A_\sigma=\{0,1\}^m$, respectively, and are decidable from $A_\sigma$.\\

Fix $j\in\{0,\ldots,m-1\}$. Define $\nu_{I,j}$ to agree with $\xi^O$ for the first $j$ bits and then use $b_\sigma$ for successive blocks of length $m$. Thus $\nu_{I,j}(\sigma\tau)=\nu_{I,j}(\sigma)b_\sigma(\tau)$ whenever $|\sigma|=j+km$ for some $k\geq0$ and $|\tau|=m$. Probabilities of shorter continuations are obtained by summing over their extensions. This defines a probability measure on $\{0,1\}^{\infty}$.\\

We verify that $\nu_{I,j}\preceq_{\mathrm{pred}}\xi^O$. At a history $\sigma v$ within a block, where $|v|<m$, consider the length-$m$ extensions of $v$. If every such extension belongs to $A_\sigma$, or if none does, use the next-bit conditionals of $\xi^O$, including when $\nu_{I,j}(\sigma v)=0$. Otherwise, both types of extension occur, so their total $b_\sigma$-probability is strictly positive, and the next-bit conditionals are obtained by dividing the appropriate finite sums of $b_\sigma$. These alternatives are decidable, and the resulting total conditionals are computable from every name of $\xi^O$. By Proposition~\ref{prop:forecast-closure}, $\nu_{I,j}\in\mathcal M^O_{\mathrm{refl}}$.\\

The construction is uniform in $m,A,l,u,j$. After a prefix-free description of $m,A,l,u$, the offset $j\in\{0,\ldots,m-1\}$ can be encoded in $\lceil\log_2m\rceil$ bits; no offset bits are needed when $m=1$. Thus, for a constant $c$ independent of these parameters,
\[
K_{\xi^O}(\nu_{I,j})\leq K(m,A,l,u)+\lceil\log_2m\rceil+c.
\]
Set
\[
B_{I,j}=\bigl(K_{\xi^O}(\nu_{I,j})+C_{\xi^O}\bigr)\ln 2
\leq\bigl(K(m,A,l,u)+\lceil\log_2m\rceil+C\bigr)\ln 2,
\]
where $C=c+C_{\xi^O}$. Let $t_i=j+1+(i-1)m$, $\sigma_i=\alpha_{<t_i}$, and $\tau_i=\alpha_{t_i:t_i+m-1}$ for $i\geq1$. Proposition~\ref{thm:forecast-relative-dominance} and the chain rule give
\[
\sum_{i\leq n}\log\frac{b_{\sigma_i}(\tau_i)}{\xi^O(\tau_i|\sigma_i)}
=\log\frac{\nu_{I,j}(\alpha_{1:j+nm})}{\xi^O(\alpha_{1:j+nm})}
\leq B_{I,j},
\]
where $\log0=-\infty$.\\

Fix $\mu\in I$. At every history $\sigma$ with $\mu(\sigma)>0$, the definition of $b_\sigma$ ensures that $b_\sigma(\tau)=0$ implies $\mu(\tau|\sigma)=0$. Consequently, $\nu_{I,j}(\sigma)>0$ whenever $\mu(\sigma)>0$. The logarithmic terms are therefore finite $\mu$-almost surely and have finite expectations for each fixed $n$.\\

Suppose $p_i=p_{\sigma_i}$, $q_i=q_{\sigma_i}$, and $r_i=\mu(A_{\sigma_i}|\sigma_i)$. Whenever $0<p_i<1$, conditioning on the history gives
\[
\mathbb E_\mu\!\left[
\left.
\log\frac{b_{\sigma_i}(\tau_i)}{\xi^O(\tau_i|\sigma_i)}
\right|
\alpha_{<t_i}
\right]
=
r_i\log\frac{q_i}{p_i}
+
(1-r_i)\log\frac{1-q_i}{1-p_i},
\]
where terms weighted by zero are taken to be zero. Since $r_i\in[l(\sigma_i),u(\sigma_i)]$ and $q_i$ is the projection of $p_i$ onto this interval, the Bernoulli calculation and Pinsker's inequality in the proof of Theorem~\ref{thm:incomplete-models} give
\[
\mathbb E_\mu\!\left[
\left.
\log\frac{b_{\sigma_i}(\tau_i)}{\xi^O(\tau_i|\sigma_i)}
\right|
\alpha_{<t_i}
\right]
\geq
2|q_i-p_i|^2.
\]
If $p_i\in\{0,1\}$, compatibility gives $r_i=q_i=p_i$, and both sides are zero. The inequality therefore holds $\mu$-almost surely in all cases.\\

Taking expectations, summing over $i\leq n$, and applying the likelihood bound yields
\[
2\sum_{i\leq n}
\mathbb E_\mu\!\left[
|q_i-p_i|^2
\right]
\leq
B_{I,j}.
\]
By the definition of $q_i$,
\[
|q_i-p_i|
=
\operatorname{dist}\!\left(
\xi^O(A_{\alpha_{<t_i}}|\alpha_{<t_i}),
[l(\alpha_{<t_i}),u(\alpha_{<t_i})]
\right).
\]
Letting $n\to\infty$, applying the monotone convergence theorem, and summing over $j\in\{0,\ldots,m-1\}$ gives
\[
2\mathbb E_\mu\!\left[
\sum_{t=1}^{\infty}
\operatorname{dist}\!\left(
\xi^O(A_{\alpha_{<t}}|\alpha_{<t}),
[l(\alpha_{<t}),u(\alpha_{<t})]
\right)^2
\right]
\leq
\sum_{j=0}^{m-1}B_{I,j},
\]
since every $t\geq1$ belongs to exactly one of these $m$ subsequences. The bound on each $B_{I,j}$ proves the stated complexity estimate. Since this non-negative series has finite expectation, it is finite $\mu$-almost surely. Hence $\operatorname{dist}(\xi^O(A_{\alpha_{<t}}|\alpha_{<t}),[l(\alpha_{<t}),u(\alpha_{<t})])\to0$ as $t\to\infty$, $\mu$-almost surely. Moreover, for every $\varepsilon>0$, the expected number of times this distance exceeds $\varepsilon$ is at most $\sum_{j=0}^{m-1}B_{I,j}/(2\varepsilon^2)$.

\end{proof}

\end{document}